\documentclass[secnumarabic, graphics,floatfix, nofootinbib,tightenlines,nobibnotes, aps, prl, 12pt]{revtex4-2}
\usepackage[english]{babel}
\usepackage[utf8]{inputenc}
\usepackage{amsfonts}
\usepackage{multirow}
\usepackage{graphicx}
\usepackage{epstopdf}
\usepackage{hyperref}
\usepackage{latexsym}
\usepackage{amsmath}
\usepackage{amssymb}

\usepackage[utf8]{inputenc}
\hypersetup{
    colorlinks=true,
    linkcolor=red,
    citecolor=blue,
}
\usepackage{color}
\usepackage[T1]{fontenc}
\usepackage{txfonts}
\usepackage{mathrsfs}
\usepackage[center]{subfigure}

\begin{document}
\setcounter{secnumdepth}{2}
\setcounter{equation}{0}

\newcommand{\bq}{\begin{equation}}
\newcommand{\eq}{\end{equation}}
\newcommand{\bqn}{\begin{eqnarray}}
\newcommand{\eqn}{\end{eqnarray}}
\newcommand{\nb}{\nonumber}
\newcommand{\lb}{\label}
\newcommand{\be}{\begin{equation}}
\newcommand{\en}{\end{equation}}
\newcommand{\PRL}{Phys. Rev. Lett.}
\newcommand{\PL}{Phys. Lett.}
\newcommand{\PR}{Phys. Rev.}
\newcommand{\CQG}{Class. Quantum Grav.}
 
\title{Two distinct types of echoes in compact objects}

\author{Shui-Fa Shen$^{1,2}$}
\author{Kai Lin$^{3,4}$}
\author{Tao Zhu$^{5}$}
\author{Yu-Peng Yan$^{6,7}$}
\author{Cheng-Gang Shao$^{8}$}
\author{Wei-Liang Qian$^{4,9}$}\email[E-mail: ]{wlqian@usp.br}

\affiliation{$^{1}$ School of intelligent manufacturing, Zhejiang Guangsha Vocational and Technical University of Construction, 322100, Jinhua, Zhejiang, China}
\affiliation{$^{2}$ School of Nuclear Science and Engineering, East China University of Technology, 330013, Nanchang, Jiangxi, China}
\affiliation{$^{3}$ Universidade Federal de Campina Grande, Campina Grande, PB, Brasil}
\affiliation{$^{4}$ Universidade de S\~ao Paulo, S\~ao Paulo, Brasil}
\affiliation{$^{5}$ Institute for Theoretical Physics and Cosmology, Zhejiang University of Technology, 310032, Hangzhou, China}
\affiliation{$^{6}$ School of Physics, Suranaree University of Technology, 30000, Nakhon Ratchasima, Thailand}
\affiliation{$^{7}$ Thailand Center of Excellence in Physics, Commission on Higher Education, 10400, Ratchathewi, Bangkok, Thailand}
\affiliation{$^{8}$ MOE Key Laboratory of Fundamental Physical Quantities Measurement, Hubei Key Laboratory of Gravitation and Quantum Physics, PGMF, and School of Physics, Huazhong University of Science and Technology, Wuhan 430074, China}
\affiliation{$^{9}$ Center for Gravitation and Cosmology, School of Physical Science and Technology, Yangzhou University, 225002, Yangzhou, Jiangsu, China}

\date{Sept. 12th, 2024}

\begin{abstract}
In the black hole perturbation theory framework, two different physical pictures for echoes in compact objects have been proposed.
The first mechanism interprets echoes as repeated reflections of gravitational waves within a potential well, where the echo period is defined by twice the distance related to the spatial displacement operator that separates two local maxima of the effective potential.
The second mechanism associates echoes with a discontinuity in the effective potential, potentially associated with specific accretion processes, without necessarily introducing a second local maximum in the effective potential.
This discontinuity leads to echo signals that are typically attenuated over time more quickly, with their period dictated by the characteristics of the transfer amplitudes.
In both scenarios, the echoes correspond to a new category of quasinormal modes with minor real parts, with their period connected to the spacing between successive modes in the frequency domain.
This work elaborates on a unified framework in compact stars that encompasses both echo mechanisms.
It suggests that these two types of echoes derive from different physical origins and can be independently triggered.
The occurrence and interplay between these two types of echoes are demonstrated through numerical simulations.
\end{abstract}

\pacs{04.60.-m; 98.80.Cq; 98.80.-k; 98.80.Bp}

\maketitle
\newpage

\section{Introduction}\label{section1}

The inception of gravitational wave (GW) observation, facilitated by direct waveform captures~\cite{agr-LIGO-01, agr-LIGO-02, agr-LIGO-03, agr-LIGO-04}, has initiated a groundbreaking phase in observational astronomy.
This novel research domain is dedicated to exploring the strong-field gravity sector, which has experienced rapid progress lately.
Notably, current endeavors in space-based laser interferometry~\cite{agr-LISA-01, agr-TianQin-01, agr-Taiji-01} have led to significant advancements aimed at achieving unmatched detector sensitivity~\cite{agr-TDI-review-01, agr-TDI-review-02, agr-TDI-Wang-03, agr-SNR-Wang-01, agr-SNR-Wang-02}.

Theoretical insights have been instrumental in assessing the feasibility of black hole spectroscopy~\cite{agr-bh-spectroscopy-05, agr-bh-spectroscopy-06, agr-bh-spectroscopy-15, agr-bh-spectroscopy-18, agr-bh-spectroscopy-20, agr-bh-spectroscopy-36}.
Specifically, in the context of real-world phenomena, sources of gravitational radiation such as black holes and neutron stars are embedded within a medium, not existing in isolation.
This consideration inevitably alters spacetime from a perfect symmetric metric, substantially modifying the GWs emitted from such an entity.
Such observations have directed research focus towards ``dirty'' black holes~\cite{agr-bh-thermodynamics-12, agr-qnm-33, agr-qnm-34, agr-qnm-54}, thus broadening the scope of black hole perturbation theory.

Within this framework, there has been a concentrated effort on constructing realistic models involving compact astrophysical bodies, including binaries of black holes or neutron stars.
The investigation of black hole quasinormal modes (QNMs)~\cite{agr-qnm-review-02, agr-qnm-review-03, agr-qnm-review-06} has been a primary focus, given their critical role during the ringdown phase post-merger.
These modes, which are dissipative oscillations reflective of the core characteristics of the black hole spacetime, adhere to several no-hair theorems~\cite{agr-bh-nohair-01, agr-bh-nohair-04}.
Initial studies by Leung \textit{et al.}~\cite{agr-qnm-33} on scalar QNMs of non-rotating ``dirty'' black holes, evaluating shifts in quasinormal frequencies through generalized logarithmic perturbation theory, laid the groundwork in this area.
Further analysis by Barausse \textit{et al.} on perturbations surrounding a central Schwarzschild black hole~\cite{agr-qnm-54} demonstrated that the resulting QNMs could exhibit significant deviations from an isolated black hole's characteristics.
Nevertheless, they posited that the influence of the astrophysical environment on black hole spectroscopy could be minimal if appropriate waveform templates were applied.
Among the scenarios examined in~\cite{agr-qnm-54}, the thin shell model was identified as having a significant effect on the QNM spectrum.
The concept of the black hole pseudospectrum, introduced by Nollert and Price~\cite{agr-qnm-35, agr-qnm-36}, aligns with these insights, showing how minimal perturbations, depicted as step functions, could remarkably influence high-overtone QNM modes, thus revealing a sizable susceptibility of the QNM spectrum to ``ultraviolet'' perturbations.
Our prior work~\cite{agr-qnm-lq-03} contended that discontinuities could alter the QNM spectrum's asymptotic behavior non-perturbatively, causing high-overtone modes to transition along the real axis rather than the imaginary frequency axis~\cite{agr-qnm-continued-fraction-12, agr-qnm-continued-fraction-23}, a phenomenon observed irrespective of the discontinuity's proximity to the horizon or its order.
Through the notion of structural stability, Jaramillo \textit{et al.}~\cite{agr-qnm-instability-07, agr-qnm-instability-13, agr-qnm-instability-14} explored the effects of metric randomizations, with their findings via the hyperboloidal coordinates~\cite{agr-qnm-hyperboloidal-03} indicating a shift of the pseudospectrum's boundary towards the real frequency axis, thereby emphasizing the inherent instability of high-overtone modes to ultraviolet perturbations.
Recent findings by Cheung \textit{et al.}~\cite{agr-qnm-instability-15} also suggest the vulnerability of even the fundamental mode to destabilization by generic perturbations.

Cardoso~\textit{et al.} introduced the concept of echoes~\cite{agr-qnm-echoes-01}, a novel phenomenon intersecting with late-stage ringing waveforms, potentially distinguishing different gravitational systems through unique near-horizon characteristics.
This concept has spurred numerous investigations~\cite{agr-qnm-echoes-40,agr-qnm-echoes-41,agr-qnm-echoes-42} into echoes across various systems, including exotic compact objects such as gravastars~\cite{agr-eco-gravastar-02, agr-eco-gravastar-03}, wormholes~\cite{agr-wormhole-01, agr-wormhole-02, agr-wormhole-10, agr-wormhole-11, agr-wormhole-12, agr-qnm-echoes-43, agr-qnm-echoes-44}, among others.

On the analytic aspect, echoes' derivation from Green's function's properties, especially its asymptotic behavior, was furthered by Mark \textit{et al.}~\cite{agr-qnm-echoes-15} through frequency domain evaluation.
Echo phenomena in compact objects were then conceptualized by reinterpreting the response waveform as a series summation of reflection and transmission amplitudes alongside the convolution integration process in the inverse Fourier transform.
From the viewpoint of a scattering process, investigations of echoes within the framework of Damour-Solodukhin type wormholes~\cite{agr-wormhole-12} by Bueno \textit{et al.}~\cite{agr-qnm-echoes-16} involved solving for specific frequencies at which the transition matrix turns singular, leading to identifiable quasinormal frequencies that give rise to echoes.
While most scenarios of echoes occur for effective potentials possessing two local maxima, some of us proposed an alternative mechanism for echoes~\cite{agr-qnm-echoes-20} without necessarily introducing a second local maximum in the effective potential. 
The Green's function's approach was adapted, and the echoes' emergence is linked to the asymptotic pole structure of the QNM spectrum~\cite{agr-qnm-lq-03, agr-qnm-lq-matrix-06} due to some minor discontinuity planted into the potential.
This suggests that discontinuity might serve an alternative ingredient for echo mechanism.

In principle, it can be argued that both approaches stand on an equivalent theoretical foundation concerning a peculiar property of the transit amplitudes.
However, from a physical perspective, they can be viewed as two distinct mechanisms for generating echoes.
In the first scenario, the effective potential is characterized by two local maxima separated by a distance, creating a potential well~\cite{agr-qnm-echoes-16, agr-qnm-echoes-22}, with Damour-Solodukhin type wormholes~\cite{agr-qnm-echoes-16} serving as a prime example.
Echoes in this context can be intuitively understood as resulting from repeated reflections of GWs within such a potential well, with the echo period mathematically determined as twice the distance between the maxima of the effective potential via the spatial displacement operator that separates the two local maxima.
The second scenario entails a degree of discontinuity within the effective potential~\cite{agr-qnm-echoes-15, agr-qnm-echoes-20}, potentially brought about by specific accretion processes giving rise to ``cuspy''~\cite{agr-dark-matter-06, agr-dark-matter-07, agr-dark-matter-08} or ``splashback''~\cite{agr-dark-matter-21, agr-dark-matter-24} profiles, without necessarily leading to a second local maximum in the effective potential.
This gives rise to echo signals typically attenuated over time, with their period dictated by the characteristics of the relevant transfer amplitudes.
For both cases, the echoes correspond to a novel category of quasinormal modes with minor real parts, and the echo period is associated with the spacing, along the direction of the real frequency axis, between successive modes~\cite{agr-qnm-echoes-20}.

The present study aims to explore this topic further, proposing a unified model within compact stars that encompasses both echo mechanisms.
We demonstrate that the two types of echoes, stemming from distinct physical origins, can be independently triggered within such a model, as evidenced through numerical simulations.
The remainder of the work is structured as follows.
In Sec.~\ref{section2}, we delve into Green's function approach for the two distinct echo mechanisms.
Subsequently, in Sec.~\ref{section3}, we provide an overview of axial gravitational perturbations in compact stars, serving as a comprehensive theoretical setup for the emergence of echoes through both mechanisms.
The master equation for the perturbations is formulated for compact stars of uniform density.
Numerical results are presented in Sec.~\ref{section4}, where spatial-temporal evolutions are calculated using the finite difference method, elucidating the principal characteristics of the echoes and the interplay between the two mechanisms.
Additional discussions and concluding remarks are given in Sec.~\ref{section5}.

\section{Two physical pictures for echoes in compact objects}\label{section2}

This section explores Green's function formalism within black hole perturbation theory and its application to echoes in compact objects.
When variable separation is achievable, the evolution of perturbations simplifies, with the dynamics predominantly governed by the radial component of the master equation~\cite{agr-qnm-review-03,agr-qnm-review-06}:
\begin{eqnarray}
\frac{\partial^2}{\partial t^2}\Psi(t, x)+\left(-\frac{\partial^2}{\partial x^2}+V_\mathrm{eff}\right)\Psi(t, x)=0 ,
\label{master_eq_ns}
\end{eqnarray}
where $x$ denotes the tortoise coordinate, and $V_\mathrm{eff}$ represents the effective potential defined by the spacetime metric, spin ${\bar{s}}$, and angular momentum $\ell$ of the waveform.
For example, the Regge-Wheeler potential $V_\mathrm{RW}$ for a Schwarzschild black hole metric is given by
\bqn
V_\mathrm{eff} = V_\mathrm{RW}=F\left[\frac{\ell(\ell+1)}{r^2}+(1-{\bar{s}}^2)\frac{r_h}{r^3}\right],
\lb{Veff_RW}
\eqn
where
\bqn
F=1-r_h/r ,
\lb{f_RW}
\eqn
$r_h=2M$ represents the horizon, with $M$ being the black hole mass, and the tortoise coordinate relates to the radial coordinate $r$ through $x=\int dr/F(r)$.

QNMs of black holes are determined by solving the eigenvalue problem presented in Eq.~\eqref{master_eq_ns} in the frequency domain:
\begin{equation}
\frac{d^2\Psi(\omega, x)}{dx^2}+[\omega_n^2-V_\mathrm{eff}(r)]\Psi(\omega, x) = 0 . \label{eq2}
\end{equation}
For asymptotically flat spacetimes, the boundary conditions are defined as:
\begin{equation}
\Psi \sim
\begin{cases}
e^{-i\omega_{n} x}, & x \to -\infty, \\
e^{+i\omega_{n} x}, & x \to +\infty,
\end{cases}
\label{master_bc0}
\end{equation}
indicating an ingoing wave at the horizon and an outgoing wave at infinity, with $n$ denoting the overtone number.
Beyond the initial burst, the waveform $\Psi$ is characterized by quasinormal oscillations with complex eigenvalues $\omega_{n}$, termed quasinormal frequencies.
The dissipative nature of the system is reflected in the imaginary components of these frequencies.

The QNM properties are intricately linked to the analytic properties of the corresponding Green's function, satisfying
\begin{equation}
\left[\frac{d^2}{dx^2}+(\omega_{n}^2-V_\mathrm{eff}(r))\right]\widetilde{G}(\omega, x,y)= \delta(x-y) .\label{DefGreen}
\end{equation}
Following established procedures~\cite{agr-qnm-21, agr-qnm-28, agr-qnm-29}, Green's function is constructed as:
\begin{equation}
\widetilde{G}(\omega, x,y)= \frac{1}{W(\omega)}f(\omega, x_<)g(\omega, x_>) ,\label{FormalGreen}
\end{equation}
where $x_<\equiv \min(x, y)$, $x_>\equiv \max(x, y)$, and
\begin{equation}
W(\omega) \equiv W(g, f) = {g} {f}' - {f} {g}' \label{DefWronskian}
\end{equation}
is the Wronskian of $f$ and $g$, the two linearly independent solutions of the corresponding homogeneous equation satisfying the boundary conditions Eq.~\eqref{master_bc0} at the horizon and infinity, respectively.

To be specific, $f$ and $g$ demonstrate the following asymptotic behaviors
\begin{equation}
f(\omega, x) \sim
\begin{cases}
e^{-i\omega x}, & x \to -\infty, \\
A_{\mathrm{out}}(\omega)e^{+i\omega x}+A_{\mathrm{in}}(\omega)e^{-i\omega x}, & x \to +\infty,
\end{cases}
\label{master_bc1}
\end{equation}
and
\begin{equation}
g(\omega, x) \sim
\begin{cases}
B_{\mathrm{out}}(\omega)e^{+i\omega x}+B_{\mathrm{in}}(\omega)e^{-i\omega x}, & x \to -\infty,\\
e^{+i\omega x}, & x \to +\infty,
\end{cases}
\label{master_bc2}
\end{equation}
in spacetimes that are asymptotically flat, remaining bounded as $t\to +\infty$ for $\Im \omega <0$.
Herein, $A_{\mathrm{in}}$, $A_{\mathrm{out}}$, $B_{\mathrm{in}}$, and $B_{\mathrm{out}}$ represent the reflection and transmission coefficients, which, although their exact expressions might not be specified in our derivations, are well-defined for a specific metric.
These waveform amplitudes adhere to the relations~\cite{book-blackhole-Frolov}
\bqn
B_{\mathrm{out}} &=& A_{\mathrm{in}},\nb\\
B_{\mathrm{in}} &=& -A_{\mathrm{out}}^*,\label{fluxConv}
\eqn
as dictated by principles of completeness and conservation of flux.
Moreover, the reflection and transmission coefficients for black holes are specified as
\bqn
\widetilde{\mathcal{R}}_\mathrm{BH}(\omega) &=& \frac{B_{\mathrm{in}}}{B_{\mathrm{out}}},\nb\\
\widetilde{\mathcal{T}}_\mathrm{BH}(\omega) &=& \frac{1}{B_{\mathrm{out}}},\label{RefTransA}
\eqn
for waves originating from $x\to -\infty$.

Echoes, akin to QNMs, are identified by the poles of Green's function.
While the inherent pole structure primarily emerges from the zeros of the Wronskian Eq.~\eqref{DefWronskian}, it may undergo alterations due to phenomena like pole skipping~\cite{adscft-pole-skipping-02, adscft-pole-skipping-05} or additional distortions introduced by an external source~\cite{agr-qnm-lq-02}.
Echoes are categorized within this context.
The discussion outlines two scenarios in which particular effective potentials lead to a unique set of poles associated with echoes in compact objects.
For the first scenario, the Wronskian is determined by segregating two potential barriers by a specified distance using a spatial displacement operator, interpreting the resultant echoes as repeated GW reflections within a potential well, with the echo period defined by twice the distance separating the two local maxima.
The second scenario incorporates discontinuity at a certain order of the effective potential, where the Wronskian is derived from the explicit forms of transmission amplitudes at the discontinuity point.
In both instances, it is demonstrated that the resulting mathematical expressions share similarities, and the echoes correspond to a new series of quasinormal modes characterized by minor real parts, with the echo period correlating to the spacing between successive modes in the frequency domain.

\subsection{The first type of echo}

In this subsection, we explore Damour-Solodukhin wormholes as a paradigm for the first type of echo.
The waves described by Eqs.~\eqref{master_bc1} and~\eqref{master_bc2} can be associated with two distinct effective potentials, $V^{(1)}$ and $V^{(2)}$, respectively.
On the right-hand side of $V^{(1)}$ and left-hand side of $V^{(2)}$, the asymptotic forms of $f$ and $g$ are as follows:
\begin{equation}
f(\omega, x) \sim A^{(1)}_{\mathrm{out}}(\omega)e^{+i\omega x}+A^{(1)}_{\mathrm{in}}(\omega)e^{-i\omega x},
\end{equation}
and
\begin{equation}
g(\omega, x) \sim B^{(2)}_{\mathrm{out}}(\omega)e^{+i\omega x}+B^{(2)}_{\mathrm{in}}(\omega)e^{-i\omega x},
\end{equation}
in an asymptotically flat spacetime.

Accounting for the potential well involves shifting the effective potential $V^{(2)}$ rightward by a distance $L$, leading to
\begin{equation}
\tilde{g}(\omega, x) \sim \tilde{B}^{(2)}_{\mathrm{out}}(\omega)e^{+i\omega x}+\tilde{B}^{(2)}_{\mathrm{in}}(\omega)e^{-i\omega x},
\end{equation}
where the transformed outgoing wave results from appropriately applying the spatial displacement operator in the frequency domain:
\bqn
\hat{P}(L)= e^{i\omega L} ,
\eqn
to the equation above. 
Specifically, it follows that
\bqn
\tilde{B}^{(2)}_{\mathrm{in}}(\omega) &=& B^{(2)}_{\mathrm{in}}(\omega) ,\nb\\
\tilde{B}^{(2)}_{\mathrm{out}}(\omega) &=& e^{2i\omega L} B^{(2)}_{\mathrm{out}}(\omega) .\label{novoB}
\eqn
Utilizing these equations, the Wronskian is found to be
\begin{equation}
W(\omega) = -i\omega\left[A^{(1)}_{\mathrm{out}}(\omega)B^{(2)}_{\mathrm{in}}(\omega)-e^{2i\omega L} A^{(1)}_{\mathrm{in}}(\omega)B^{(2)}_{\mathrm{out}}(\omega)\right]. \label{WronskianDS}
\end{equation}

The emergence of echoes is inferred from the zeros of the obtained Wronskian, which are poles of the Green's function. 
Assuming the original effective potentials $V^{(1)}$ and $V^{(2)}$ do not inherently produce echoes, the frequency $\omega$'s amplitudes $A^{(1)}_{\mathrm{in}}(\omega)$, $A^{(1)}_{\mathrm{out}}(\omega)$, $B^{(2)}_{\mathrm{in}}(\omega)$, and $B^{(2)}_{\mathrm{out}}(\omega)$ are moderate.
In regions far from their zeros and poles, they are considered constant.
This premise leads to a new sequence of zeros for the Wronskian, corresponding to poles in the Green's function, distributed evenly and parallel to the real frequency axis.
Under the above apprximation, if $\omega$ is a root, then $\omega+n\pi/L$ (for any integer $n$) is also a root, forming a branch of the QNM spectrum.
The impact of these poles on the time-domain Green's function results in echoes.
The spacing between successive poles set at $2L$, and the proximity of this branch to the real axis define, respectively, the echo period and observational viability.

\subsection{The second type of echo}\label{secII2}

Now, we turn our attention to the second type of echo.
In this case, the emergence of echoes is closely associated with the presence of a discontinuity in the effective potential.
The discontinuity that triggers these echoes may be positioned near what would traditionally be considered the horizon or at a more distant location relative to the compact object.
The core principles underlying the formulation of these echoes remain consistent across different physical systems.
As discussed previously, the mathematical framework for deriving echoes from exotic compact objects has been established in~\cite{agr-qnm-echoes-15, agr-qnm-echoes-20}.

For simplicity, let us consider the discontinuity to be near the horizon.
For such a case, the ingoing wave is slightly altered by a fraction of the outgoing wave, while the latter remains unaffected.
Specifically, we describe this interaction as
\bqn
\tilde{f}(\omega,x) &=& f(\omega,x)+\widetilde{\mathcal{C}}(\omega)g(\omega,x) ,\label{h1Cform}
\eqn
representing a mixture of the ingoing and outgoing waves, as characterized by the asymptotic behaviors delineated in Eqs.~\eqref{master_bc1} and~\eqref{master_bc2}.

Assuming the vicinity of the discontinuity, denoted by $x_c$, features a significantly suppressed effective potential $V=V_c\sim 0$, the ingoing wave $f$ can be approximated as a superposition of two plane waves
\bqn
\tilde{f}(\omega,x)\propto e^{-i\omega (x-x_c)} +\widetilde{\mathcal{R}}(\omega)e^{i\omega (x-x_c)} ,\label{h1Rform}
\eqn
where the reflection amplitude $\widetilde{\mathcal{R}}$ is primarily determined by the unique characteristics of the compact object.

The relationship between $\widetilde{\mathcal{C}}(\omega)$ and $\widetilde{\mathcal{R}}(\omega)$ is given by
\bqn
\widetilde{\mathcal{C}}(\omega) = \frac{e^{-2i\omega x_c}\widetilde{\mathcal{T}}_\mathrm{BH}(\omega)\widetilde{\mathcal{R}}(\omega)}{1-e^{-2i\omega x_c}\widetilde{\mathcal{R}}_\mathrm{BH}(\omega)\widetilde{\mathcal{R}}(\omega)} ,\label{relCR}
\eqn
with the reflection and transmission amplitudes defined by Eq.~\eqref{RefTransA}.
The derivation of Eq.~\eqref{relCR} involves aligning the coefficients of the wave functions at an arbitrary asymptotical point where their expressions converge to plane waves, detailed further in Appx.~\ref{App1}.

Consequently, the frequency domain Green's function, as given by Eq.~\eqref{FormalGreen}, is computed to be
\bqn
\widetilde{G}(\omega,x,y)
&=& \frac{\tilde{f}(\omega,y) g(\omega,x)}{W(g,\tilde{f})}\nb\\
&=& \frac{f(\omega,y) g(\omega,x)}{W(g,f)}+\widetilde{\mathcal{C}}(\omega)\frac{g(\omega,y) g(\omega,x)}{W(g,f)}\nb\\
&\equiv& \widetilde{G}_\mathrm{BH} +\widetilde{\mathcal{C}}(\omega)\frac{g(\omega,y) g(\omega,x)}{W_\mathrm{BH}} ,\label{Gtilde_h3}
\eqn
where $W_\mathrm{BH}=W(g,f)$ denotes the Wronskian, and $\widetilde{G}_\mathrm{BH}$ is the Green's function corresponding to the original black hole metric.
This formulation suggests that the QNMs inherent to the black hole largely persist within the wormhole metric, with echoes attributed to the poles in $\widetilde{\mathcal{C}}(\omega)$ as defined by Eq.~\eqref{relCR}.

Once more, the echoes can be analyzed by assessing the poles of the above Green's function.
In this framework, the manifestation of echoes is linked to the phase shift embedded in $\widetilde{\mathcal{C}}(\omega)$~\cite{agr-qnm-echoes-15, agr-qnm-echoes-20, agr-strong-lensing-correlator-15}, which explicitly depends on the location of discontinuity $x_c$.
Again, as the original black hole metric does not imply any echo phenomenon, the quantities $\widetilde{\mathcal{T}}_\mathrm{BH}$ and $\widetilde{\mathcal{R}}_\mathrm{BH}$ are moderate functions of the frequency $\omega$ and will be treated as constants when analyzing the echo modes.
For a simplified scenario where the discontinuity is implemented by truncating the effective potential at $x_c$ from the inside, the reflection amplitude is found to possess the following form
\bqn
\widetilde{\mathcal{R}}(\omega)= \frac{\sqrt{\omega^2-V_c}-\omega}{\sqrt{\omega^2-V_c}+\omega} ,\label{RForm2echo}
\eqn
using the WKB approximation, whose derivation is also relegated to Appx.~\ref{App1}.
Using Eq.~\eqref{relCR}, it is readily verified that for asymptotical modes $\omega \gg 1$, we have
\bqn
\widetilde{\mathcal{C}}(\omega) \sim \frac{1}{2\omega}e^{-2i\omega x_c}\widetilde{\mathcal{T}}_\mathrm{BH}(\omega)V_c \ll 1 .
\eqn
The succeeding arguments are mainly reminiscent of the preceding subsection.
Specifically, if $\omega$ is a pole of Eq.~\eqref{relCR} and therefore the Green's function Eq.~\eqref{Gtilde_h3}, $\omega+n\pi/x_c$ (where $n$ is an arbitrary integer) is also a pole.
The union of such poles gives rise to an additional branch of QNMs, contributing to echoes of a period of $2x_c$\footnote{Rather than representing an absolute coordinate, we note that $x_c$
essentially measures a relative distance. It quantifies the offset between the location of the effective potential, relative to which the transmission coefficients are defined and evaluated, and the discontinuity. This has been implied in the preceding derivations.}.

Before closing this section, we observe a nuanced distinction between the two scenarios.
The first scenario does not inherently involve any discontinuity within the metric.
Here, the echo period is solely determined by the separation of two local maxima within an essentially continuous effective potential.
In our derivation, we have assumed that the waveforms within the region spanning these maxima are asymptotic in the sense that they can be expressed as linear combinations of plane waves.
This assumption is crucial for the mathematical assessment of the Wronskian.
Conversely, the second scenario does not necessitate the presence of a secondary maximum within the effective potential.
Instead, it posits that the discontinuity is minor and sits far from the potential's peak, permitting only a negligible portion of the plane wave to move in reverse to merge into the initially outgoing wave.
In this case, the echoes hinge critically on the singularity of the coefficient of the added waveform, whose assessment generally requires some form of model-based assumption or simplification.
Nevertheless, in both scenarios, the emergent QNM spectrum stems from a term of the form $e^{2i\omega L}$, under the presumption that the other terms are moderate frequency functions, notably lacking any exponential dependency on $\omega$.
These scenarios are elucidated by examining the zeros in the Wronskian Eq.~\eqref{WronskianDS} or the denominator of Eq.~\eqref{relCR}.
Specifically, in the first scenario, the pertinent contribution derives from altering the transit amplitude as indicated in the second line of Eq.~\eqref{novoB}, as a result of the translation operation.
This scenario suggests echoes result from the GWs ricocheting between the two local maxima of the effective potential.
In contrast, the second scenario's contribution is linked to the singularity within a minor fraction of the outgoing wave, as delineated by Eqs.~\eqref{h1Cform} and~\eqref{relCR}, underscoring the distinct physical implications of the discontinuity introduced.
As indicated in Sec.~\ref{section4}, echoes emanating from such conditions are usually more substantially attenuated over time.

\section{The master equation in the compact star of uniform density}\label{section3}

The above two types of echoes might be present simultaneously in the context of compact objects.
This is because the effective potential of a compact star may possess two local maxima, while the star's surface usually introduces some discontinuity.
In the literature, the well-known w-modes~\cite{agr-qnm-star-07, agr-qnm-star-09, agr-qnm-star-23, agr-qnm-star-26} are readily recognized as echoes in star quasinormal modes.
However, to our knowledge, an explicit discussion regarding the nuance difference elaborated in the foregoing section has yet to be carried out.

In what follows, we derive the master equation for axial gravitational perturbations in stars of uniform density.
To enhance the reader's understanding, we adopt a minimalist approach that focuses on a spherically symmetric compact star with uniform density, making the physical content more accessible. 
The spacetime metric possess the form
\bqn
\lb{1}
ds^2=-h(r)dt^2+\frac{dr^2}{f(r)}+r^2\left(d\theta^2+\sin^2\theta d\varphi^2\right) .
\eqn

On the outside of the star $r\ge r_b$, where $r_b$ is the radial coordinate of the star's surface, the metric is essentially Schwarzschild, which reads
\bqn
\lb{2}
f(r)=h(r)=1-\frac{2M_S}{r} ,
\eqn
where $M_S$ is the mass of the compact object.
The tortoise coordinate reads $x\equiv r_*=\int dr/\sqrt{fh}=r+r_S\log\left(\frac{r}{r_S}-1\right)$.

Beneath the surface of a star $r\le r_b$ with uniform density $\rho=\frac{3M_S}{4\pi r_b^3}$, one has~\cite{book-general-relativity-Weinberg}
\bqn
\lb{3}
f(r)&=&1-\frac{2M(r)}{r}\nb ,\\
h(r)&=&\frac{1}{4}\left(3\sqrt{1-\frac{2M(r)}{r}}-\sqrt{1-\frac{2M(r)r^2}{r_b^3}}\right)^2\nb ,\\
M(r)&=&\int^r_04\pi r'^2 \rho dr'=M_S\frac{r^3}{r_b^3}\nb ,\\
p(r)&=&-\rho\left(\frac{\sqrt{1-\frac{2M(r)}{r}}-\sqrt{1-\frac{2M(r)r^2}{r_b^3}}}{3\sqrt{1-\frac{2M(r)}{r}}-\sqrt{1-\frac{2M(r)r^2}{r_b^3}}}\right) .
\eqn
For the tortoise coordinate, the constant of integration is chosen so that $x(r=0)=0$.

By using the method of separation of variables, the axial gravitational perturbations are governed by~\cite{agr-qnm-star-30}
\bqn
\lb{4}
&&\frac{\partial^2\Psi}{\partial r^2}-\frac{\partial^2\Psi}{\partial t^2}-V(r)\Psi=0\nb ,\\
&&V_\text{star}(r)=\frac{h}{r^3}\left(\ell(\ell+1)r+4\pi(\rho-p)r^3-6M\right)\nb ,\\
&&V_\text{BH}(r)=\frac{h}{r^3}\left(\ell(\ell+1)r-6M_S\right) ,
\eqn
where $L$ is the angular momentum.
At spatial infinity, the boundary condition dictates that the waveform $\Psi$ is an asymptotic outgoing wave.
At the star's center, it must be regular with vanishing flux.
Moreover, the junction conditions at the star's surface reads
\bqn
\lb{5}
\Psi_\text{inside}(r_b)&=&\Psi_\text{outside}(r_b)\nb ,\\
\partial_r\Psi_\text{inside}(r_b)&=&\partial_r\Psi_\text{outside}(r_b) ,
\eqn
or
\bqn
\partial_x\Psi_\text{inside}(r_b)=\partial_x\Psi_\text{outside}(r_b) .
\eqn

\section{Numerical results about the emergence of two types of echoes}\label{section4}

\begin{figure*}[tbp]
\centering
\includegraphics[width=0.4\columnwidth]{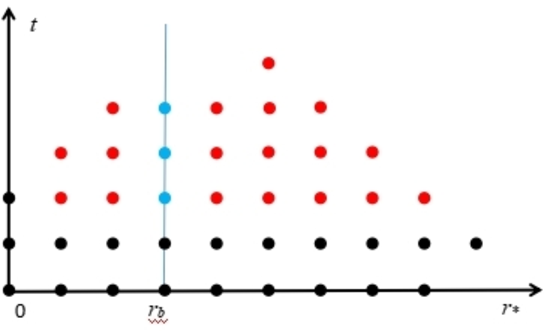}
\caption{The grid layouts of finite difference method employed in the present study.
As described in the text, the calculations are implemented in $(r_*\equiv x, t)$ coordinates.
The black points indicate the grids to which the initial and boundary conditions are assigned. 
The finite difference method uses Eqs.~\eqref{masDes} and~\eqref{surStar} to evaluate the remaining grids in red.}
\lb{FDM_scheme}
\end{figure*}

The numerical calculations will be performed using the finite difference method~\cite{agr-qnm-finite-difference-01, agr-qnm-finite-difference-02, agr-qnm-finite-difference-03}.
We will adopt the scheme recently proposed by some of us~\cite{agr-qnm-lq-09}.
In the literature, most implementations are performed in Eddington-Finkelstein coordinates $(u=t-x$, $v=t+x)$, where one implements the initial condition on $v=0$ and the boundary condition on $u=0$, 
In such circumstances, in the far region where the effective potential asymptotically approaches zero, the initial perturbations will never attain the $u$ axis due to causality constraints.
However, in the region where the effective potential does not vanish, there is a chance that the speed of signal propagation exceeds the unit, and initial perturbations placed on the $u$ axis might traverse the $v$ axis.
In other words, the boundary condition enforced for $u=0$ might interfere with the initial perturbations' free propagation.

Conversely, in our proposed scheme, the calculations are implemented in the $(x, t)$ coordinates, and the initial condition is placed on the $t=0$ slice.
As a result, the concerns regarding causality will no longer be relevant.
In particular, we adopt the initial conditions $\Psi(t=0)$ of a Gaussian form and $\partial_t\Psi(t=0)$.
One proceeds to discretize the spacetime coordinates and approximates the partial derivatives by first-order finite differences.
To be specific, we denote 
\bqn
t_i&=& t_0+i\Delta t , \nb\\
x_{j}&=& x_0+j\Delta x 
\eqn
and 
\bqn
\Psi^i_j &=& \Psi(t=t_i,x=x_{j}) , \nb\\
V_j &=& V(x=x_{j}) ,
\eqn
and therefore, the master equation Eq.~\eqref{4} becomes
\bqn
\lb{masDes}
\Psi^{i+1}_j = -\Psi^{i-1}_j+\frac{\Delta t^2}{\Delta x^2}\left(\Psi^i_{j-1}+\Psi^{i-1}_j\right)
+\left(2-2\frac{\Delta t^2}{\Delta x^2}-\Delta t^2V_j\right)\Psi^i_j .
\eqn

The junction condition Eq.~\eqref{5} reads
\bqn
\frac{\Psi^i_{j_b}-\Psi^i_{j_b-1}}{\Delta x}=\frac{\Psi^i_{j_b+1}-\Psi^i_{j_b}}{\Delta x} , \lb{surStar0}
\eqn
where one uses the subscript $b$ to denote the grid on the star's surface.
Eq.~\eqref{surStar0} can be simplified to give
\bqn
\lb{surStar}
\Psi^i_{j_b}=\frac{1}{2}\left(\Psi^i_{j_b+1}+\Psi^i_{j_b-1}\right) ,
\eqn
where the subscript $b$ indicates that the grid is on the star's surface.

As shown in Fig.~\ref{FDM_scheme}, the numerical iterative process can be performed by using Eqs.~\eqref{masDes} and~\eqref{surStar}.
Given the black grids on the boundary, the temporal evolution is implemented by inferring the values of the red grids.
For most grids, Eq.~\eqref{masDes} is utilized to determine the grid values for the next time step, except for those on the star's surface.
The latter is determined by employing Eq.~\eqref{surStar}.
To avoid the von Neumann instability~\cite{agr-qnm-finite-difference-04, agr-qnm-lq-01}, we choose 
\bqn
\frac{\Delta t^2}{\Delta x^2}+\frac{\Delta t^2}{4}V_\text{max}<1 .
\eqn

\begin{figure*}[tbp]
\centering
\includegraphics[width=0.4\columnwidth]{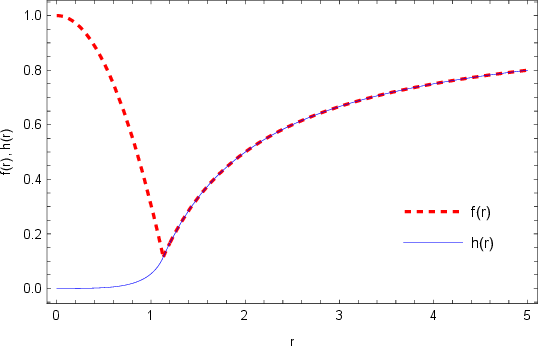}
\includegraphics[width=0.4\columnwidth]{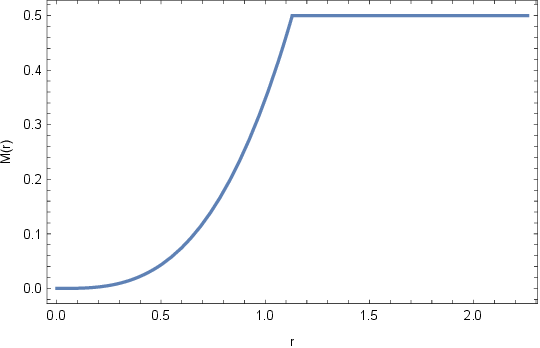}\\
\includegraphics[width=0.4\columnwidth]{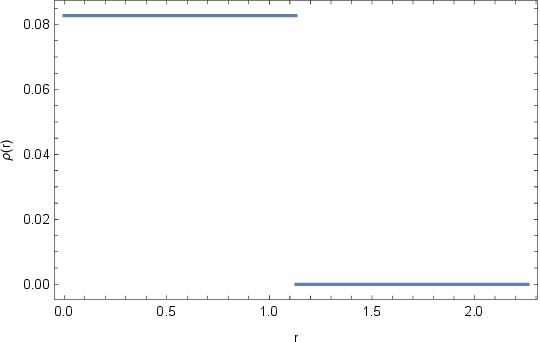}
\includegraphics[width=0.4\columnwidth]{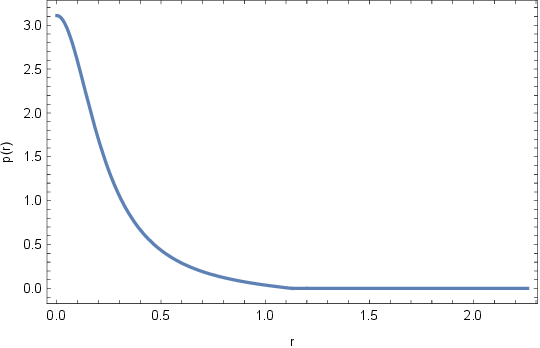}\\
\caption{The metric functions $f(r)$ and $h(r)$ (top-left), the mass function $M(r)$ (top-right), density $\rho(r)$ (bottom-left), and pressure $p(r)$ (bottom-right) of the compact star in question shown as a function of the radial coordinate $r$.
The plots are obtained using the metric parameters $M_S=\frac12$ and $r_b=1.13$.}
\lb{Metric_star}
\end{figure*}

\begin{figure*}[tbp]
\centering
\includegraphics[width=0.4\columnwidth]{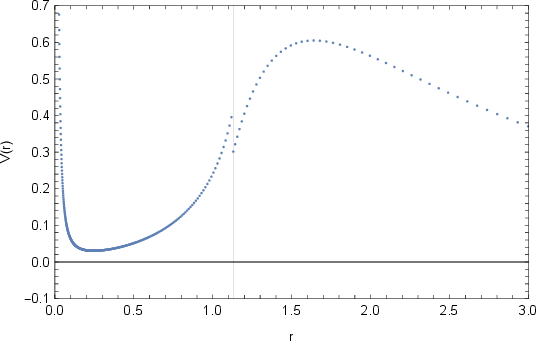}
\includegraphics[width=0.4\columnwidth]{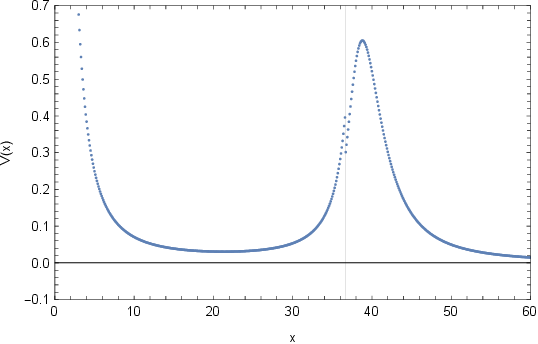}\\
\includegraphics[width=0.4\columnwidth]{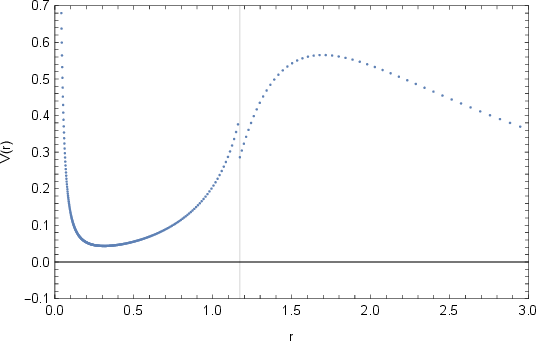}
\includegraphics[width=0.4\columnwidth]{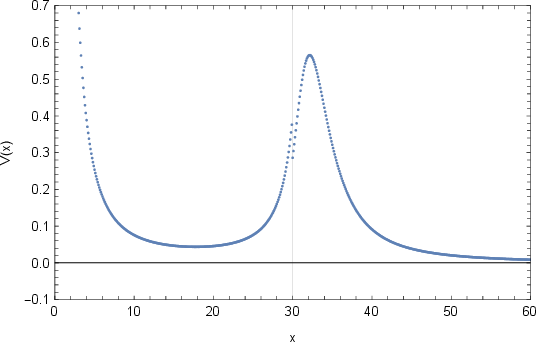}\\
\caption{The effective potentials of the compact star as a function of the radial coordinate $r$ (left column) and tortoise coordinate $x$ (right column).
The top row is obtained using the metric parameters $M_S=\frac12$, $r_b=1.13$, and $\ell=2$, while the bottom row is evaluated using $M_S=\frac12$, $r_b=1.14$, and $\ell=2$.
For both cases, the effective potential features two local maxima at $r=0$ and $r=1.65$ while a discontinuity is also present at $r_b$.
In the tortoise coordinates, the distances between the two maxima for the two sets of parameters are $L=40$ (top) and $32$ (bottom).}
\lb{V_star}
\end{figure*}

Now, we proceed to present the numerical results.
For the first types of echoes, we elaborate two sets of parameters: $M_S=\frac12$, $r_b=1.13$, $\ell=2$, and $M_S=\frac12$, $r_b=1.14$, $\ell=2$.
The metric functions are presented in Fig.~\ref{Metric_star}, and the corresponding effective potentials are shown in Fig.~\ref{V_star}.
When presented in radial coordinate $r$, the forms of the effective potentials are primarily identical, particularly the positions of the two maxima.
However, the difference is apparent in terms of the tortoise coordinate $x$.
For both parameters, echoes are demonstrated, as shown in Fig.~\ref{V_echoes}.
The spatial-temporal evolution indicates that initial perturbations propagate at $v=1$ and are reflected at both maxima.
The temporal evolution clearly shows the occurrence of echoes.
The distance between the two maxima governs the resulting echoes' period.
Even though a discontinuity is present, it is irrelevant to the observed echoes.

\begin{figure*}[tbp]
\centering
\includegraphics[width=0.4\columnwidth]{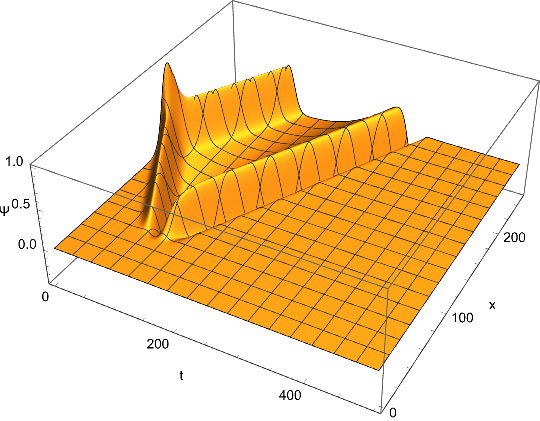}
\includegraphics[width=0.4\columnwidth]{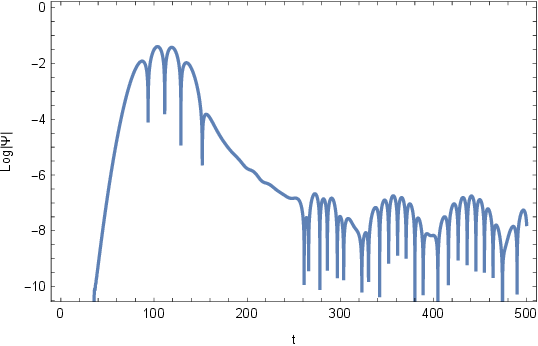}\\
\includegraphics[width=0.4\columnwidth]{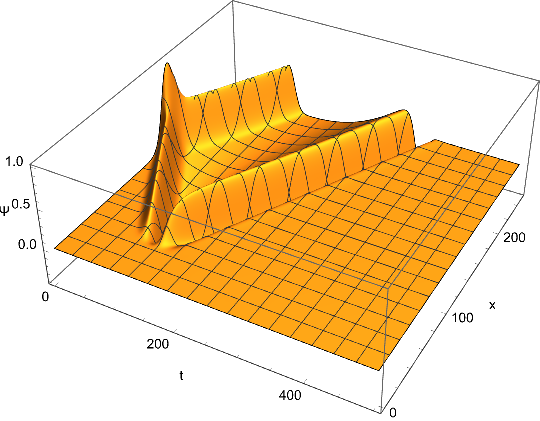}
\includegraphics[width=0.4\columnwidth]{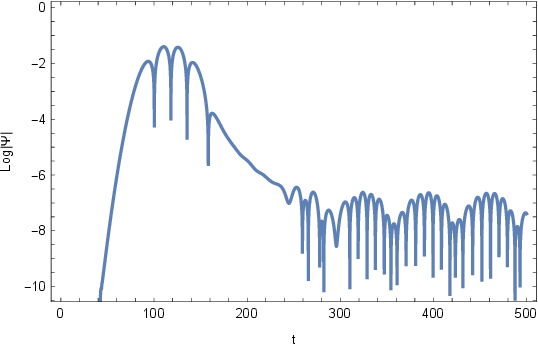}\\
\caption{The echoes for the two sets of metric parameters whose effective potentials are given in Fig.~\ref{V_star}.
The spatial-temporal evoluations are shown in the left column, while the temporal evolutions are presented in the right column.
The echo periods are found to be $T \simeq 80$ (top) and $64$ (bottom), in accordance with twice the distance between the two maxima, while irrelevant to the position of the discontinuity.}
\lb{V_echoes}
\end{figure*}

We consider the metric parameters $M_S=\frac12$, $r_b=100$, and $\ell=2$ to demonstrate the second type of echo.
As shown in Fig.~\ref{V_star_p3}, since the discontinuity is placed beyond the would-be maximum of Regge-Wheeler effective potential, the resulting effective potential does not feature a second maximum.
Moreover, the discontinuity is visually insignificant.
The resulting echoes are presented in Fig.~\ref{V_echoes_p3}.
The echoes' period is found to be $T \simeq 200$, in accordance with twice the distance between the maximum and the discontinuity of the effective potential.
In contrast to Fig.~\ref{V_echoes}, the magnitude of the echoes is suppressed in time.
This explains why such echoes were not observed in Fig.~\ref{V_echoes}, as it is likely buried inside the echoes formed by the first type.

\begin{figure*}[tbp]
\centering
\includegraphics[width=0.4\columnwidth]{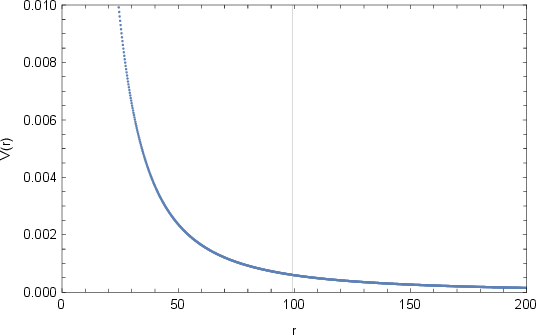}
\includegraphics[width=0.4\columnwidth]{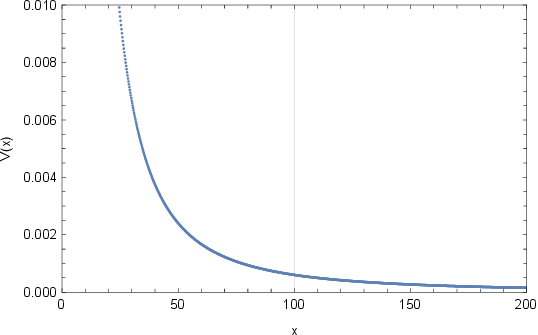}\\
\caption{The effective potentials of the compact star as a function of the radial coordinate $r$ (left) and tortoise coordinate $x$ (right).
The results are obtained using the metric parameters $M_S=\frac12$, $r_b=100$, and $\ell=2$.
The effective potential does not possess any local maximum but features a discontinuity at the star's surface.
In both coordinates, it is apparent that the discontinuity is visually insignificant.}
\lb{V_star_p3}
\end{figure*}

\begin{figure*}[tbp]
\centering
\includegraphics[width=0.4\columnwidth]{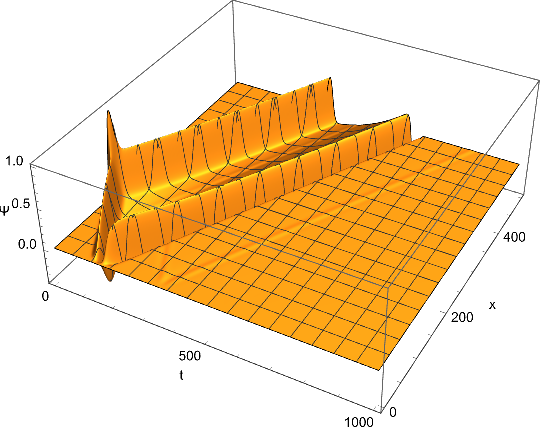}
\includegraphics[width=0.4\columnwidth]{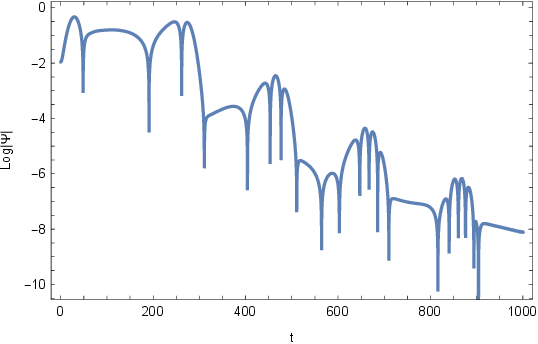}\\
\caption{The echoes for the third set of metric parameters whose effective potentials are given in Fig.~\ref{V_star_p3}.
The spatial-temporal and temporal evoluations are shown in the left and right panels.
The echoes' period is found to be $T \simeq 200$, in accordance with twice the distance between the maximum and the discontinuity of the effective potential.}
\lb{V_echoes_p3}
\end{figure*}

\section{Concluding remarks}\label{section5}

Echoes, emerging as a novel observable in the GW spectrum, offer a unique window into the physics of compact objects and their environments. 
By exploring the nuances of echo mechanisms through a unified model, this study contributes to a deeper understanding of the complex interplay between GWs and the structure of compact objects. 
Through numerical simulations, we demonstrate the possibility of independent triggering of distinct echo types, highlighting the diversity of physical processes that can manifest in the GW signals from compact stars.
In particular, for the simplified model adopted in the present study, the more attenuated echoes might actually occur earlier.
Therefore, it might still be observed if the first type of echo also exists and will be more persistent during evolution.
Regarding the ongoing spaceborne GW detection programs, the findings presented herein not only shed light on the theoretical underpinnings of echo phenomena but also pave the way for future observational strategies to detect and interpret echoes in GW data.

For a given total mass, the results presented in Sec.~\ref{section4} indicate that the second type of echo becomes relevant only when the star's size is significant.
Specifically, the criterion for the emergence of the second type of echo is that the star's surface lies outside the {\it would-have-been} local maximum of the Regge-Wheeler effective potential, i.e., $r_b \gtrapprox r_\mathrm{max}$, where $r_\mathrm{max} = 3M$ in the eikonal limit.
At the same time, one might pursue other observational channels.
The above criterion ensures that such an object will likely not possess a photon sphere, specifically $r_p = 3M$ for the Regge-Wheeler potential, resulting in a distinct optical image compared to objects emitting the first type of echo.
From an empirical perspective, we argue that this condition is feasible in realistic scenarios, and various compact objects may exhibit the second type of echo.
For instance, neutron stars typically have a radius of 10-15 kilometers, while their Schwarzschild radius is about 3 kilometers~\cite{agr-neutron-star-20}.
This implies that neutron stars usually do not have a photon sphere and are subject to the second type of echo.
On the other hand, as the density of the compact object increases further, it might become optically ``dark'' due to the presence of a photon sphere while exhibiting the first type of echo.
Furthermore, beyond the specific quasinormal oscillations of stars discussed in this work, we speculate that the second type of echo signal could have broader relevance in a universal context.
For instance, on a larger scale, discontinuities arising from phenomena like Bondi accretion around compact objects could give rise to such echoes, which might be used to probe their presence.
Similarly, these echoes might manifest in the gravitational wave spectrum of compact binaries, as surfaces and resulting discontinuities are inherent features of any compact object.
In conclusion, our findings regarding the two types of echoes in compact stars enrich the theoretical framework for investigating the universe's most extreme gravitational environments.
This potentially opens new pathways for testing the predictions of general relativity and exploring alternative theories of gravity, contributing to our ongoing quest to understand the cosmos.

\section*{acknowledgments}

We also gratefully acknowledge the financial support from
Funda\c{c}\~ao de Amparo \`a Pesquisa do Estado de S\~ao Paulo (FAPESP),
Funda\c{c}\~ao de Amparo \`a Pesquisa do Estado do Rio de Janeiro (FAPERJ),
Conselho Nacional de Desenvolvimento Cient\'{\i}fico e Tecnol\'ogico (CNPq),
Coordena\c{c}\~ao de Aperfei\c{c}oamento de Pessoal de N\'ivel Superior (CAPES),
A part of this work was developed under the project Institutos Nacionais de Ci\^{e}ncias e Tecnologia - F\'isica Nuclear e Aplica\c{c}\~{o}es (INCT/FNA) Proc. No. 464898/2014-5.
This research is also supported by the Center for Scientific Computing (NCC/GridUNESP) of S\~ao Paulo State University (UNESP). T. Zhu is supported by the National Key Research and Development Program of China Grant No. 2020YFC2201503, and the Zhejiang Provincial Natural Science Foundation of China under Grant No. LR21A050001 and No. LY20A050002, the National Natural Science Foundation of China under Grant No. 12275238, and the Fundamental Research Funds for the Provincial Universities of Zhejiang in China under Grant No. RF-A2019015.

\appendix

\section{The reflection amplitude at the point discontinuity}\label{App1}

In this appendix, we derive the specific forms of the transit amplitudes utilized in Eqs.~\eqref{relCR} and~\eqref{RForm2echo} in the main text.

To derive Eq.~\eqref{relCR}, one substitutes the asymptotic forms given by the first rows of Eqs.~\eqref{master_bc1} and~\eqref{master_bc2} into Eq.~\eqref{h1Cform}.
Subsequently, one matches the coefficients of the wave functions $e^{i\omega x}$ and $e^{-i\omega x}$ between Eqs.~\eqref{h1Cform} and~\eqref{h1Rform} and finds
\bqn
\frac{1+\widetilde{\mathcal{C}}(\omega)B_\mathrm{in}}{e^{i\omega x_c}}=\frac{\widetilde{\mathcal{C}}(\omega)B_\mathrm{out}}{\widetilde{\mathcal{R}}(\omega)e^{-i\omega x_c}} ,
\eqn
which can be simplified to give
\bqn
\widetilde{\mathcal{C}}(\omega) = \frac{e^{-2i\omega x_c}\widetilde{\mathcal{T}}\mathrm{BH}(\omega)\widetilde{\mathcal{R}}(\omega)}{1-e^{-2i\omega x_c}\widetilde{\mathcal{R}}\mathrm{BH}(\omega)\widetilde{\mathcal{R}}(\omega)} ,
\eqn
where the reflection and transmission amplitudes are defined by Eq.~\eqref{RefTransA}.

To derive Eq.~\eqref{RForm2echo}, one utilizes the WKB approximation and the waveform on the r.h.s. of the truncation point $x\ge x_c$ is given by
\bqn
\tilde{f}(\omega, x)=C e^{iS(x_0,x)} + D e^{-iS(x_0,x)},
\eqn
where, at the lowest-order approximation that suffices for our case, we have
\bqn
S(x_0, x)=\int_{x_0}^x k(x')dx',
\eqn
where $k(x)=\sqrt{\omega^2-V(x)}$, and $x_0$ can be taken somewhat arbitrarily in the region where the WKB formula is relevant.
Without loss of generosity, we assume $x_0=0$.
While for $x \le x_c$, we have
\bqn
f = e^{-i\omega x} .
\eqn

The Wronskian vanishes at the point of discontinuity $x=x_c$, which gives
\bqn
\left.iS'(x_0,x)\frac{C e^{iS(0,x)} - D e^{-iS(0,x)}}{C e^{iS(0,x)} + D e^{-iS(0,x)}}\right|_{x=x_c} = -i\omega ,
\eqn
By taking into account $S'=k(x)$ and $V(x)=V_c\sim 0$, we can approximate
\bqn
\frac{C}{D}=\frac{k(x)-\omega}{k(x)+\omega}e^{-2iS(0,x_c)}
=\frac{\sqrt{\omega^2-V_c}-\omega}{\sqrt{\omega^2-V_c}+\omega}e^{-2i\omega x_c} .
\eqn
By comparing against the form of Eq.~\eqref{h1Cform}, one finds
\bqn
\widetilde{\mathcal{R}}(\omega) = e^{2i\omega x_c}\frac{C}{D} = \frac{\sqrt{\omega^2-V_c}-\omega}{\sqrt{\omega^2-V_c}+\omega},
\eqn
which is the result used in Eq.~\eqref{RForm2echo}.

\bibliographystyle{h-physrev}
\bibliography{references_qian}

\end{document}